# Multi-band metasurface-driven surface-enhanced infrared absorption spectroscopy for improved characterization of in-situ electrochemical reactions

*Malo Duportal[1], Luca M. Berger[2], Stefan A. Maier[3,4,2], Andreas Tittl[2,\*], Katharina Krischer[1,\*]*

[1] Department of Physics, Technical University of Munich, 85748 Garching, Germany
[2] Chair in Hybrid Nanosystems, Nanoinstitute Munich, and Center for NanoScience, Faculty of Physics, Ludwig-Maximilians-University Munich, Königinstraße 10, 80539 München, Germany
[3] School of Physics and Astronomy, Monash University, Wellington Rd, Clayton VIC 3800, Australia
[4] Department of Physics, Imperial College London, SW7 2AZ London, United Kingdom
\* E-mail address: Andreas.Tittl@physik.uni-muenchen.de; krischer@tum.de

**Abstract**

Surface-enhanced spectroscopy techniques are the method-of-choice to characterize adsorbed intermediates occurring during electrochemical reactions, which are crucial in realizing a green sustainable future. Characterizing species with low coverages or short lifetimes have so far been limited by low signal enhancement. Recently, metasurface-driven surface-enhanced infrared absorption spectroscopy (SEIRAS) has been pioneered as a promising narrowband technology to study single vibrational modes of electrochemical interfaces during CO oxidation. However, many reactions involve several species or configurations of adsorption that need to be monitored simultaneously requiring reproducible and broadband sensing platforms to provide a clear understanding of the underlying electrochemical processes. Here, we experimentally realize multi-band metasurface-driven SEIRAS for the in-situ study of electrochemical $CO_2$ reduction on a Pt surface. We develop an easily reproducible and spectrally-tunable platinum nano-slot metasurface. Two CO adsorption configurations at 2030 cm$^{-1}$ and 1840 cm$^{-1}$ are locally enhanced as a proof of concept that can be extended to more vibrational bands. Our platform provides a 41-fold enhancement in the detection of characteristic absorption signals compared to conventional broadband electrochemically roughened platinum films. A straightforward methodology is outlined starting by baselining our system in CO saturated environment and clearly detecting both configurations of adsorption, in particular the hitherto hardly detectable CO bridge configuration. Then, thanks to the signal enhancement provided by our platform, we find that the CO bridge configuration on platinum does not play a significant role during $CO_2$ reduction in an alkaline environment. We anticipate that our technology will guide researchers in developing similar sensing platforms to detect challenging intermediates, low surface coverages, and species with short lifetimes.

1. **Introduction**

The study of intermediates occurring during electrochemical reactions remains a challenge due to low surface coverages and short lifetimes. Raman and IR spectroscopy are powerful *in-situ* optical characterization techniques that can detect the rotational or vibrational modes of molecules. During the CO$_2$ reduction reaction (CO$_2$RR) a key intermediate is CO, which is detectable with IR spectroscopic techniques.[1] However, as IR characterization techniques suffer from high losses by aqueous electrolyte, two geometries are typically used to acquire IR spectra of molecules adsorbed at electrode surfaces. Firstly, the external reflectance approach consists in squeezing a thin layer of electrolyte between the working electrode and a prism to reduce the optical losses.[2,3] The advantage of the external reflectance approach is a high degree of freedom in choosing the morphology and thickness of the studied material. However, although the electrolyte layer is thin it still suffers from dampened signals due to light absorption from the electrolyte. Moreover, due to the requirement of a thin electrolyte layer diffusional processes are strongly limited, making the external reflectance approach an inadequate technique for *in situ* studies of electrocatalytic reactions, such as the CO$_2$RR.

A second geometry used to decrease optical losses from the electrolyte is to operate a sensing platform in an attenuated total internal reflection (ATR) configuration.[2–4] As light is totally reflected on a surface, evanescent waves with an exponential decay probe the other side of the reflective surface. Due to the evanescent approach, the losses stemming from the electrolyte are minimized. This so-called Kretchmann configuration leaves the electrode freely accessible for transport processes to and from the electrolyte. However, it requires thin film electrodes as the quickly decaying evanescent waves must be able to penetrate the electrolyte past the metal/electrolyte interface to probe the analyte. But, as the metal layer is decreased in thickness the electrochemical stability decreases.[3] In addition, ATR infrared absorption spectroscopy at plane electrodes yields only low IR signals. This disadvantage is overcome when rough film electrodes are prepared, e.g. by electrochemical roughening. Then, the electron density of nanoparticles with a linear dimension of the order of the wavelength of IR irradiation can come into resonance with the electromagnetic wave, leading to an enhanced vibrational signature.[5] The latter methods is usually referred to as ATR surface enhanced infrared absorption spectroscopy, or, in short as ATR-SEIRAS. However, with ATR-SEIRAS besides the stability of the film electrodes, reproducibility of the acquired spectra becomes an issue. The intensity of the vibrational bands strongly depends on the distribution of the nanoparticle size resulting in an uncontrolled and random signal enhancement making this technique difficult to reproduce and often unreliable.[6]

Recently, a promising nanophotonic-electrochemical platform based on ATR-SEIRAS was designed employing a nanostructured platinum surface to characterize CO during CO oxidation.[7] The precisely nanostructured platinum metasurface that integrated SEIRAS with cyclic voltammetry for the study of electrochemical interfaces provided a clear and reproducible method to produce SEIRAS-active electrodes. Specifically, the electric near-field enhancement produced by the platinum-based nano-slot metasurface was shown to amplify the *in-situ* generated signal traces of the vibrational mode of linearly adsorbed CO ($CO_{linear}$) at 2033 cm$^{-1}$. Changes in the reflection intensity based on the coupling of the metasurface-driven resonances and the vibrational modes of the adsorbed species controllably enhanced their characteristic signal traces. However, the reported nanophotonic-electrochemical platform only featured one resonance and could therefore only spectrally target and enhance one molecular vibrational mode.

Here, we develop multi-band metasurface-driven SEIRAS for the improved characterization of *in situ* electrochemical reactions (Figure 1a). We demonstrate its successful operation during the $CO_2RR$ where the molecular signals of two configurations of adsorption of CO on platinum are resolved: on-top ($CO_{linear}$) and bridged bound ($CO_{bridge}$) CO molecules, respectively (Figure 1b). So far, little attention has been paid to the involvement of bridge site configurations of CO on Pt during the $CO_2RR$. The few studies that have been conducted focused on acidic media.[8–11] A fundamental and systematic study of bridge site configurations during the $CO_2RR$ in alkaline media is still missing. We use a platinum nano-slot metasurface on a $CaF_2$ substrate featuring two arrays that were each numerically modeled and tuned to spectrally target one of the two aforementioned characteristic molecular vibrations of the $CO_2RR$. For each vibrational mode, a unique array with a spectrally targeted resonance was fabricated. Each array locally enhanced the electric near fields and enhanced the corresponding molecular signal traces. Our multiband approach can be extended to multiple vibrational bands.

SEIRAS was performed in an ATR geometry (Figure 1c) to maintain free accessibility of the electrode surface and minimize the contribution of the electrolyte to the IR spectrum. We validated the nanophotonic-electrochemical platform by following the oxidation of CO into $CO_2$ during a cathodic polarization, measuring simultaneously the top and bridge site adsorption of CO. The clear detection and enhancement of the top and bridge adsorption configurations of CO on Pt were confirmed by observing the typical Stark shift in the molecular signal traces. Then, we followed the reduction of CO₂ *in situ* to study the same adsorption sites. The contrast between the CO saturated signal traces compared to those measured during the CO$_2$RR provided conclusive insights into the adsorption characteristics of intermediates emerging during the CO$_2$RR. Our results suggest that the CO$_2$RR in alkaline environments proceeds *via* CO molecules adsorbed mainly as CO$_{linear}$ and not as CO$_{bridge}$. Finally, we established a methodology to implement similar multi-band nanophotonic-electrochemical platforms, providing a framework for future research in this area.

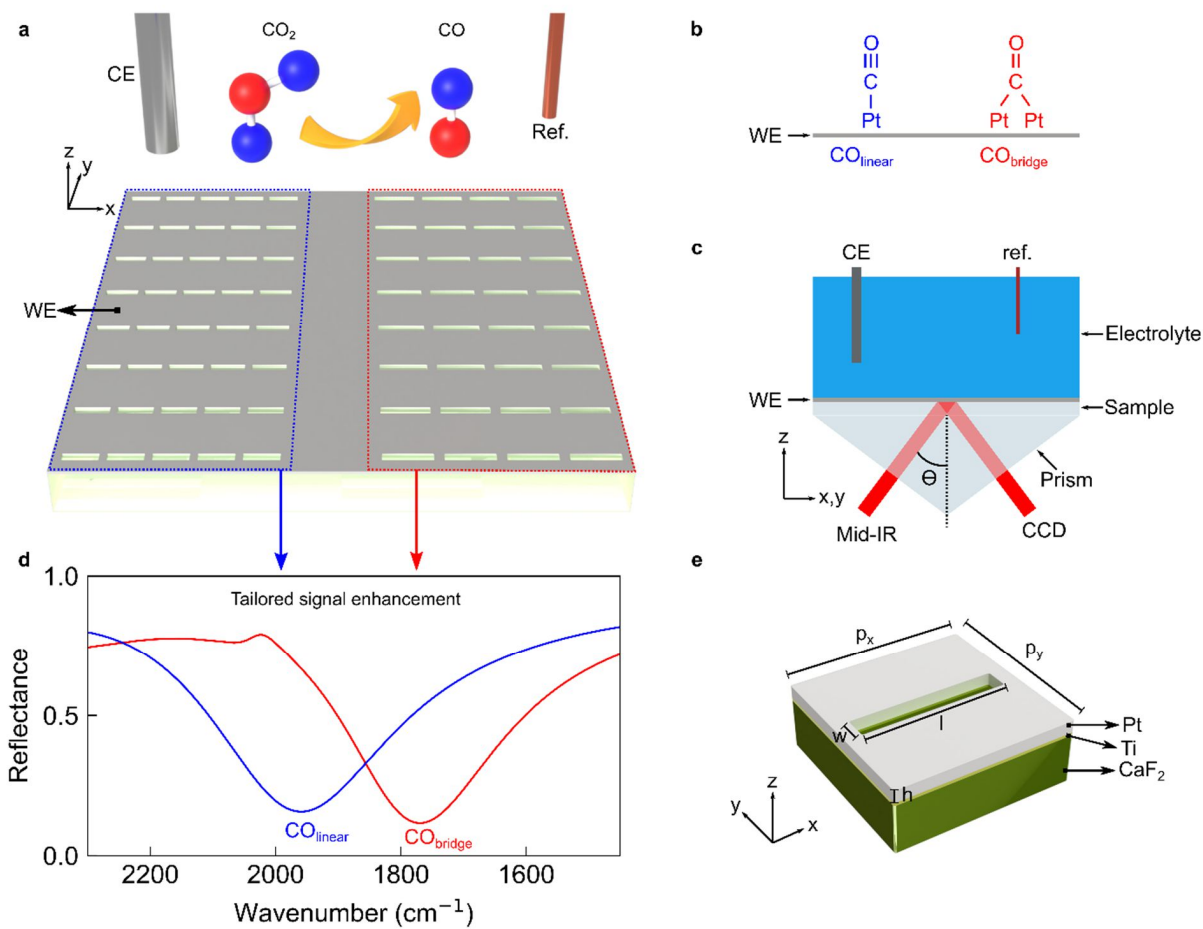

**Figure 1.** Concept and numerical design of the multi-band nanophotonic-electrochemical platform. (a) Schematic of the multi-band platinum-based nano-slot metasurface to study *in-situ* CO₂ reduction. (b) Schematic showing the chemical structure of the two adsorption configurations of CO on platinum. (c) Schematic illustrating the metasurface in an electrochemical chamber filled with electrolyte that was illuminated from below in an ATR geometry. (d) The resonances

stemming from two metasurface arrays on Pt. The shared geometrical parameters were $h$=30 nm, $w$=180 nm, $p_y$=1420 nm, $px$–$l$=230 nm with the slot length being swept from (blue) $l$=1370 nm to (red) $l$=1580. (e) Sketch of the unit cell. A 1 nm thick Ti adhesion layer was utilized in the structure's fabrication for improved adhesion of Pt on CaF$_2$.

## 2. Experimental Section

*Numerical Simulations.*

The numerical simulations in this study utilized CST Studio Suite (2021) which uses the finite-element frequency-domain Maxwell solver. For the simulation of CaF$_2$, a refractive index of 1.4 was used, while the surrounding medium was represented by water with a refractive index of 1.33. Platinum was modeled using its experimental complex refractive index data.[12] To introduce linearly polarized light at an angle of incidence of 72° into the system, an impedance-matched open port with a perfectly matched layer was employed. Since light experiences total internal reflection at this angle at the CaF$_2$-Pt interface, the boundary opposite the open port was set as a perfect electric conductor. The unit cell was defined and simulated as an infinite periodic array using Floquet boundaries.

*Analytical analysis of resonances.*

The resonances were characterized in terms of their radiative ($\gamma_{rad}$) and intrinsic ($\gamma_{int}$) damping rates, from which their total Q-factor could be determined as $Q_{tot} = \frac{\nu_0}{2(\gamma_{rad}+\gamma_{int})}$, where $\nu_0$ is the central wavenumber. We employed temporal coupled mode theory[13] according to Ref.[14] describing a resonator with a single port that supported reflected waves and a single resonance that coupled to the far-field *via* the coupling constant $\kappa = \sqrt{2\gamma_{rad}}$. Additionally, an intrinsic loss channel introduced damping to the resonance at a rate $\gamma_{int}$. The reflectance spectra $R$ were fitted by

$$R = 1 - \frac{4\gamma_{rad}\gamma_{int}}{(\nu - \nu_0)^2 + (\gamma_{rad} + \gamma_{int})^2},$$

where $\nu$ is the wavenumber.

*Multi-band Metasurface Fabrication.*

The multi-band metasurface fabrication was similar to the protocol provided in Ref.[7] Instead of one array, two arrays were fabricated ca. 500 μm apart to fit within the window of the focal plane array of the IR-spectrometer. The arrays were each ca. 3×2 mm in size showing that large arrays could be made. The arrays had to be large as the Fourier transform IR spectroscopy device did not have and did not need a focusing microscope objective. We chose $CaF_2$ as the substrate due to its transparent properties within the mid-IR spectral range, as well as its high chemical stability and low solubility. Before the experiments, the substrate underwent a thorough cleaning process, involving an acetone bath in an ultrasonic bath, followed by oxygen plasma treatment. Subsequently, the substrate was spin-coated with an adhesion promoter (Surpass 4000), followed by a layer of negative-tone photoresist (ma-N 2403). The photoresist was baked at 100 °C for 60 s, and a conducting layer (ESpacer 300Z) was deposited using spin coating. The metasurface patterns were generated by defining a unit cell and replicating it in both the *x* and *y* directions. Electron-beam lithography (Raith Eline Plus) was employed to write the patterns, utilizing an acceleration voltage of 30 kV and an aperture of 20 μm. The exposed resist was developed in ma-D 525 for 70 s at room temperature. Thereafter, a titanium adhesion layer (1 nm at 0.4 Å $s^{-1}$) and a platinum film (30 nm at 2 Å $s^{-1}$) were deposited on the patterned surface using electron-beam evaporation (PRO Line PVD 75, Lesker). Finally, the fabrication process was completed with an overnight lift-off process in mr-REM 700. For the *in-situ* SEIRAS measurements, a pure 30 nm thick platinum film on a 1 nm titanium layer on $CaF_2$ was utilized as a reference.

*Surface-Enhanced Infrared Absorption Spectroscopy Measurements.*

SEIRAS was conducted using a Vertex 80 spectrometer equipped with an IMAC Focal Plane Array macro imaging accessory from Bruker. A specular reflection unit (VeeMax III from PIKE Technologies) paired with a $CaF_2$ prism and light polarizer introduced light at 72° w.r.t. an electrochemical jackfish cell, thereby enabling the attenuated total internal reflection geometry. A focal plane array detector (64 × 64 MCT detectors) was used to characterize the optical properties of the multi-band nano-slot metasurface. By integrating the measured (absorbance) signal across the spectral range corresponding to the metasurface-driven resonances (1600-2800 $cm^{-1}$) a pixelated two-dimensional heat map of the sample was created to identify the pixels corresponding to the nanostructured arrays (Figure 2a). Then, the pixels corresponding to each array were averaged to construct the final spectra. The samples were cleaned *via* electrochemical cycling. Prior to each measurement, an initial background was acquired using p-polarized light. The metasurface-driven resonances were measured *in situ* using s-polarized light. Each spectrum was acquired at a resolution of 4 cm$^{-1}$ and by averaging 10 scans. The data was treated by applying a baseline correction and Savitzky-Golay filter. The enhancement provided by the multiband metasurface was determined by comparing the area of the peaks associated with the vibrational modes of the CO$_{linear}$ and CO$_{bridge}$ configurations to the corresponding measurement performed on an unstructured Pt film (30 nm).

*Electrochemical Measurements.*

A classical three-electrode system was implemented by using the Pt multi-band metasurface as the working electrode, a Pt wire as counter electrode and a saturated calomel electrode (E = 0.244 V$_{SHE}$) as reference electrode. The electrolyte was a 0.5 mol.L$^{-1}$ K$_2$CO$_3$ solution (pH 11.9), saturated with either Ar, CO, or CO$_2$. Before the first characterization, the cleanliness of the electrode surface was verified by performing a cyclic voltammogram at a scan rate of 20 mV s$^{-1}$ in Ar-saturated electrolyte.

The behavior of the nano-slots was characterized in CO-saturated electrolyte. After bubbling CO for 2 hours, cyclic voltammetry was performed from +1650 mV$_{RHE}$ to -85 mV$_{RHE}$, using a scan rate of 0.25 mV.s$^{-1}$.

Finally, we saturated the electrolyte with CO$_2$, which decreased the pH to approximately 8. After 2h of gas bubbling, cyclic voltammograms were measured with a slow scan rate of 0.25 mV s$^{-1}$, from the OCP to + 1425 mV$_{RHE}$ and then back to +25 mV$_{RHE}$. SEIRAS spectra were acquired in intervals of 100 mV.

## 3. Results and Discussion

*Numerical Design of Catalytic Multi-Band Nano-Slot Metasurface.*

We start the implementation of our catalytic multi-band nano-slot metasurface by defining the fundamental unit cell for the numerical simulations consisting of a solitary slot within a continuous platinum film immersed in water on $CaF_2$ (**Figure 1e**). The geometrical parameters of the unit cell can be changed to tune the resonance strength and position of the metasurface. As the goal of our investigations was multi-band signal enhancement, the unit cell parameters were separately tuned to two resonance frequencies matching the vibration frequencies of two different adsorption configurations of CO on platinum. Specifically, the goal was a straightforward approach to create two adjacent nanostructured arrays on platinum that would each enhance one of the two vibrational frequencies.

A common approach to achieving this is to scale all geometrical dimensions of the system at the same time, used for example in biosensing[15] and catalysis[16]. However, constraints in the fabrication with negative resists limited the unit cell length in *y* to a minimum of 1.4 µm due to proximity effects. Proximity effects arise due to the scattering of electrons in the resist and substrate due to exposure of an electron beam.[17] We found that the slots merged and the quality of the lift-off procedure decreased as the unit cell length in *y* was decreased below 1.4 µm. Therefore, we simplified the resonance tuning protocol allowing only the slot length to change and taking into account a shift of ca. 80 cm$^{-1}$ between simulation and experiment. By increasing the slot length from 1370 nm to 1580 nm while leaving all other parameters unchanged the metasurface-driven resonance was redshifted by ca. 150 cm$^{-1}$ (Figure 1d), corresponding to the spectral separation of the vibrational modes.

The Rayleigh anomaly was tuned to the smaller frequency side of the resonance to ensure optimal sensing performance. To improve the characterization of the resonances, the fitting model was adapted specifically for total internal reflection (see Experimental section). The Q-factor and coupling ratio $\gamma_e/\gamma_i$ were obtained by fitting the simulated resonance in reflectance using temporal coupled mode theory (see Experimental Section). Based on our simulations, the multi-band nano-slot metasurface achieved a modulation of ca. 84% and 88% in reflection, a Q-factor of 4.3 and 5.0, and a ratio of external to intrinsic coupling of 2.6 and 2.1 for the higher and lower frequency resonances, respectively. Our system can be further optimized by maximizing the modulation in reflection or absorption to push it toward its critical coupling condition.

*Multi-band Metasurface Characterization.*

According to the literature, the vibrational modes of the $CO_{linear}$ and $CO_{bridge}$ are expected to occur at ca. 2050 cm$^{-1}$ [18–22] and 1850 cm$^{-1}$ [18,19,21,23], respectively. First, we characterized the optical properties of our multi-band metasurface in electrolyte saturated with Ar and CO using SEIRAS. An example of the heat map obtained by integrating the IR spectra collected by the focal plane array detector is provided in Figure 2a. The two pink-colored areas correspond to the two metasurface-arrays designed to enhance CO detection. The quality of the fabricated slots was verified by scanning electron microscopy images (Figure 2b and 2c). The signal received by the high-frequency array (Figure 2a, bottom array) was averaged, resulting in a resonance spectrally positioned at ca. 2030 cm$^{-1}$ in the Ar saturated electrolyte (Figure 2d). On the other hand, the average signal from the low-frequency array (Figure 2a, right array) produced a resonance located at ca. 1860 cm$^{-1}$ (Figure 2e). In both cases, the system presents a near critically-coupled behavior between the metasurface-driven resonances and the vibrational modes of CO,[7] leading to a dip in the high and low-frequency resonances at 2046 cm$^{-1}$ and 1848 cm$^{-1}$, respectively. To extract the signal of the $CO_{linear}$ and $CO_{bridge}$ vibrational modes more clearly, then, the reflectance spectra with, *R*, and without, $R_0$, adsorbed CO were converted into their differential absorbance, $-\ln\frac{R}{R_0}$, to separate the metasurface-driven resonance from the CO signals (Figure 2f and 2g). Furthermore, a comparison of the differential absorbance in the regions of the $CO_{linear}$ and $CO_{bridge}$, obtained on an unstructured Pt film (30 nm thick) and with our multi-band nano-slot metasurface, indicates that the high-frequency array exhibited a signal enhancement of 41. This signal enhancement is higher than the enhancement reported in our previous work,[7] attributed to an improved liftoff procedure. The signal enhancement provided by the low-frequency array cannot be reliably estimated, as the $CO_{bridge}$ signal was not clearly distinguishable on the unstructured Pt film. The $CO_{bridge}$ signal could only be observed here with the multi-band metasurface due to its high signal enhancing properties.

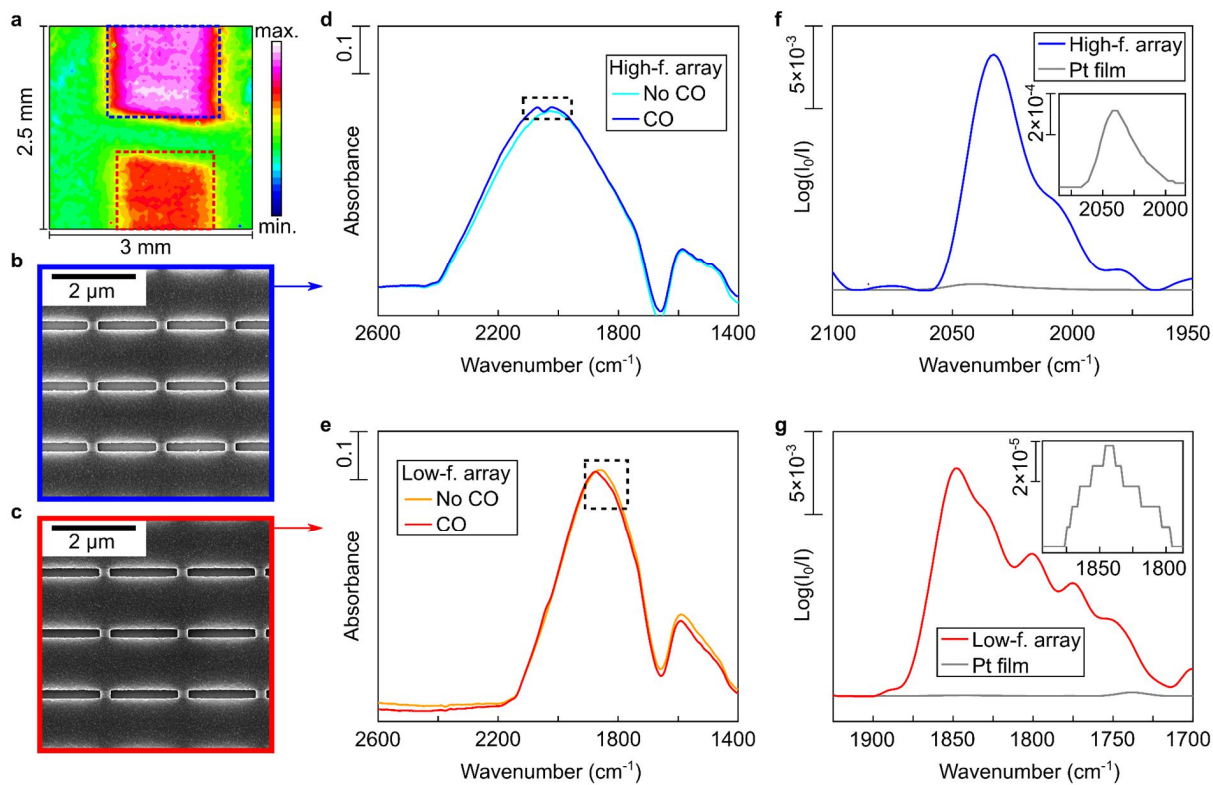

**Figure 2.** Characterization of the multi-band nano-slot metasurface. (a) Heat map of the metasurface obtained by integrating the SEIRAS signal from 1600 to 2800 cm$^{-1}$. (b-c) SEM pictures of the (b) high and (c) low-frequency arrays, indicated in (a) *via* blue and red dashed boxes, respectively. (d-e) Resonances with and without CO for the (d) high and (e) low-frequency arrays in 0.5M K$_2$CO$_3$. (f-g) Comparison of the differential absorbance obtained with the (f) high and (g) low-frequency array with an unstructured Pt film (30 nm thick, inset) and with the multi-band nano-slot metasurface.

*Behavior in CO saturated electrolyte.*

As a proof-of-concept, the behavior of the multi-band nano-slot metasurface in 0.5M K$_2$CO$_3$ saturated with CO is provided here by using electrochemical voltammetry. A cathodic scan was conducted with a rate of 0.25 mV.s$^{-1}$ starting at 1650 mV$_{RHE}$, as shown in Figure 3a. As already observed in the literature,[7,24,25] the increase in the current around 950 mV$_{RHE}$ followed by a plateau corresponds to the oxidation of CO into CO$_2$, which is limited by CO diffusion from the bulk of the electrolyte to the Pt surface. Then, the following decrease in current at 350 mV$_{RHE}$ indicates the end of the region where CO was oxidized. Finally, the slight drop in current around 0 mV$_{RHE}$ can be attributed to the hydrogen evolution reaction. SEIRAS spectra were acquired at intervals of 100 mV during a cathodic scan. The potential regions in which CO was or was not detected are shown as red and blue regions, respectively.

The evolution of the IR spectrum with the electrical polarization of the high (Figure 3b) and low-frequency (Figure 3c) arrays shows the successful detection of the CO$_{linear}$ and CO$_{bridge}$, respectively. The vibrational modes appeared as peaks in the differential absorbance spectra. Regarding the behavior of the CO$_{linear}$, a spectral shift of approximately 63 cm$^{-1}$.V$^{-1}$ was observed (Figure 3d) from 450 mV$_{RHE}$ to -50 mV$_{RHE}$, which is in line with the literature[7,26–29] and can be attributed to either a higher π-back-donation from the metal to CO[27,30] and/or to the Stark effect.[28,30,31] Additionally, a significant spectral redshift attributed to the decrease of the dipole-dipole interactions as the coverage decreases[32,33] was resolved from 650 to 450 mV$_{RHE}$ (Figure 3d), thanks to the high resolution achieved with our platform. Concerning the behavior of the CO$_{bridge}$ on the low-frequency array, a distinct peak was resolved exhibiting a similar behavior as the CO$_{linear}$. However, the observed Stark shift in this case was smaller, resulting in approximately 21 cm$^{-1}$.V$^{-1}$. This difference in Stark shift between CO$_{linear}$ and CO$_{bridge}$ has been observed in some other literature.[28,29] However, the origin of this difference is still controversial as other authors suggested that both configurations should provide the same Stark shift.[34,35] Our observations can be explained by the smaller IR cross-section of the CO$_{bridge}$ compared to the CO$_{linear}$.[34] When CO is adsorbed linearly on the platinum surface, it strongly binds to the Pt atoms,[36] which could explain the larger Stark shift. Furthermore, the electric field from the metal surface could have affected the CO$_{linear}$ more significantly, leading to a larger change in its vibrational frequency. On the other hand, the configuration of the CO$_{bridge}$ may have resulted in a weaker interaction with the metal surface.[36] In that case, the electric field from the metal surface could have had a smaller impact on this configuration, resulting in a smaller Stark shift.

At higher potentials, the redshift with decreasing coverage was observed next to a broadening of the peak from 550-750 mV$_{RHE}$, which needs further investigation. A Fano-type asymmetric peak was observed at the same position as the CO$_{linear}$ due to the off-resonance coupling between the resonance of the low-frequency array and the vibrational mode of the CO$_{linear}$.[37] Interestingly, the area of the CO$_{linear}$ peak slightly decreased from 650mV$_{RHE}$ to -50mV$_{RHE}$, while the area of the CO$_{bridge}$ peak showed a continuous increase (Figure 3e). The area and intensity of the CO peaks are related to the coverage of the CO$_{linear}$ and CO$_{bridge}$, suggesting a transfer from a CO$_{linear}$ to a CO$_{bridge}$ configuration as the cathodic potential increased. According to some authors,[29,36] the competition between CO and hydrogen adsorption on Pt in the cathodic region could have been responsible for this transition. Furthermore, the barrier for CO diffusion from a top site to a bridge site was theoretically predicted to be very small.[38] These findings can explain the increase in the area of the peak attributed to the CO$_{bridge}$ to the detriment of the CO$_{linear}$ peak area, which decreases.

To conclude, our multi-band nano-slot metasurface selectively and simultaneously enhanced and detected with a high accuracy the behavior of the CO$_{linear}$ and CO$_{bridge}$. In the next section, the study focused on the behavior of adsorbed CO during the CO$_2$ reduction reaction.

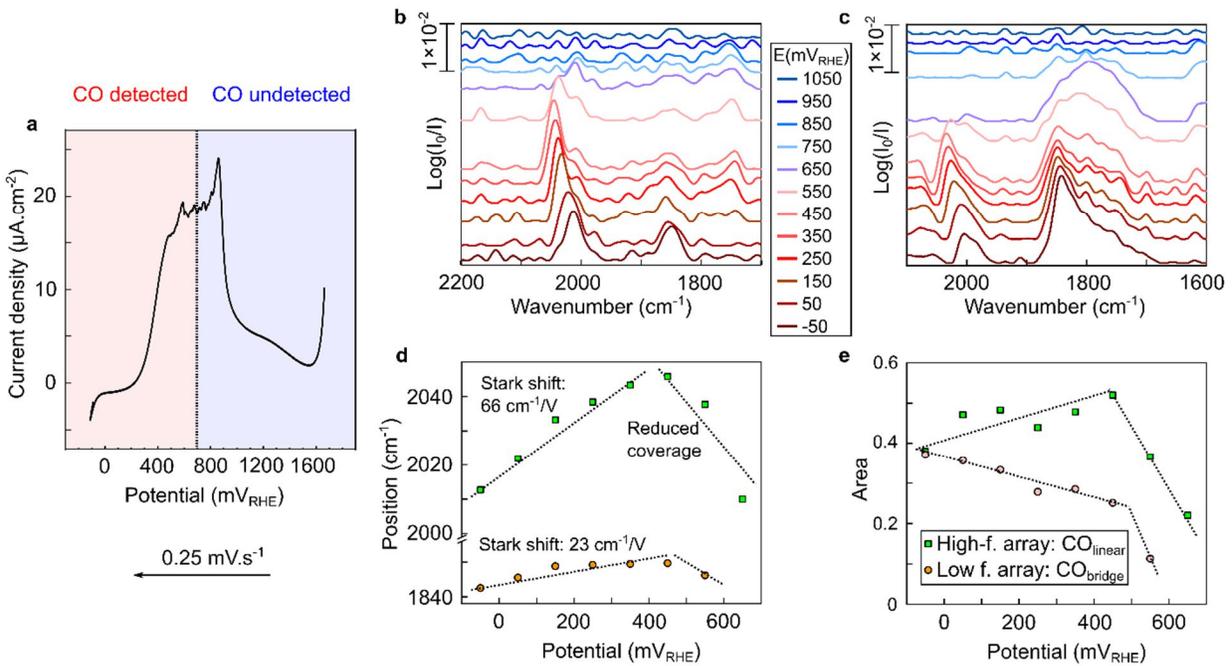

**Figure 3.** Cathodic polarization of the multi-band metasurface in 0.5M K$_2$CO$_3$ saturated with CO. (a) Evolution of the current density during the cathodic polarization at 0.25 mV.s$^{-1}$. (b-c) Evolution of the IR spectra with the potential acquired by the arrays optimized for the (b) CO$_{linear}$ and (c) CO$_{bridge}$ detection. (d-e) Evolution of the (d) position and (e) area of peaks with the potential.

*Reduction of CO$_2$.*

The cathodic scan of the cyclic voltammetry of the multi-band metasurface in 0.5M K$_2$CO$_3$ saturated with CO$_2$ shows a dip around 400 mV$_{RHE}$ (Figure 4a), which could be attributed to the chemisorption of hydrogen.[39–41] A second peak appeared around 250 mV$_{RHE}$, which is attributed to CO$_2$ reduction.[22] The drop in the current starting at 0 mV$_{RHE}$ is due to the hydrogen evolution reaction and the positive current peak observed after reversing the scan direction stems from the hydrogen oxidation reaction.[28] Finally, a small oxidative peak was observed around 550 mV$_{RHE}$ and is attributed to the oxidation of the previously formed CO.[42] Looking at the IR spectra of the high-frequency array optimized for the CO$_{linear}$ during the cathodic scan (Figure 4b), a peak was observed around 200 mV$_{RHE}$. This peak indicates the presence of adsorbed CO, which is directly correlated to the reduction peak observed in the voltammogram. Moreover, this peak became more pronounced and more defined with higher cathodic polarizations.

Moving to the low-frequency array optimized for the CO$_{bridge}$ detection, at the highest applied cathodic polarization (0 mV$_{RHE}$) a small peak became discernible at around 1785 cm$^{-1}$ in the IR spectra (Figure 4c). This peak is hardly above the detection limit despite the high resolution achieved with our multi-band nanophotonic-electrochemical platform and appeared at significantly higher cathodic potentials (200 mV difference) than the CO$_{linear}$ peak obtained with the high-frequency array. Our findings suggest that the CO$_{bridge}$ is not significantly involved in the CO$_2$ reduction process on Pt. Some authors[19,43] have suggested that the favorable configuration of adsorption of CO on Pt is the CO$_{linear}$. Moreover, the literature suggests that the CO$_{bridge}$ formation occurs once the CO coverage approaches its maximum limit where a transfer from the CO$_{linear}$ to the CO$_{bridge}$ configuration takes place.[29,36]

During the anodic scan, the CO$_{linear}$ peak maintained a constant amplitude (Figure 4d) but exhibited a classic Stark shift between 0 to 400 mV$_{RHE}$, followed by an intensity decrease and a small redshift attributed to reduced coverage around 500 mV$_{RHE}$. Then, the peak disappeared, which can be directly correlated with the oxidation peak observed in the voltammogramm (Figure 4a). In contrast, the peak attributed to the CO$_{bridge}$ disappeared at 100 mV$_{RHE}$ (Figure 4e), supporting our conclusion that the CO$_{bridge}$ is not significantly involved in the CO$_2$ reduction on platinum in an alkaline environment. The distortion of the baseline at around 2000 cm$^{-1}$ is attributed to the strong off-resonance coupling between the metasurface-driven resonance of the low-frequency array and the vibrational mode of $CO_{linear}$.[37] Thanks to the high resolution and signal enhancing properties provided by our nanophotonic-electrochemical platform, Figure 4 suggests that the $CO_{bridge}$ is not significantly involved in the reduction of $CO_2$ in alkaline electrolyte. This is noteworthy because the $CO_{bridge}$ configuration was detected in acidic electrolyte during the $CO_2RR$.[8–11] The difference in the CO adsorption behavior between alkaline and acidic media could be due to a modification of the competition with the hydrogen evolution reaction. In fact, in an alkaline environment, the Heyrovsky reaction is often considered the dominant pathway, which differs from the Volmer-Tafel mechanism observed in acidic media.[44–46] The Heyrovsky reaction involves the formation of hydroxide ions near the Pt surface, which could interact with adsorbed $CO_2$ and CO to favor the formation of the $CO_{linear}$. These considerations are in agreement with our observation that $CO_{bridge}$ did not significantly participate in the $CO_2RR$ in $K_2CO_3$, i.e., in an alkaline environment.

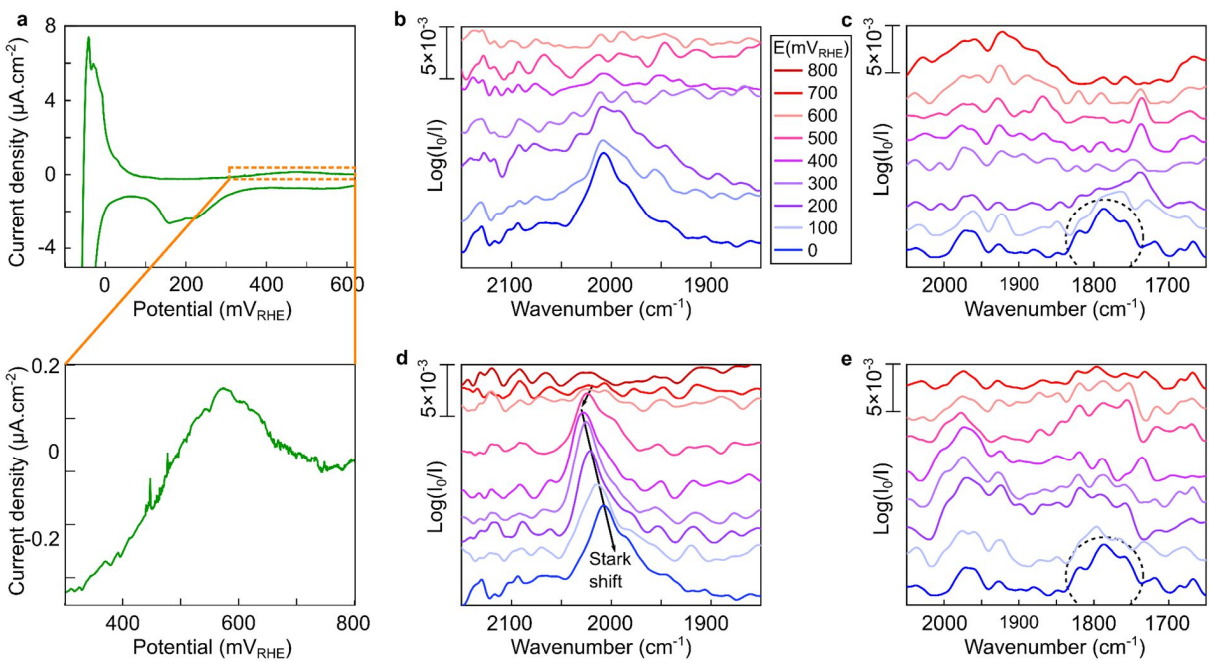

**Figure 4.** Cyclic voltammetry of the multi-band metasurface in 0.5M $K_2CO_3$ saturated with $CO_2$. (a) Evolution of the current density during the polarization at 0.25 mV.s$^{-1}$. A zoom-out of the current density is shown. (b-c) Evolution of IR spectra with the potential acquired on the (b) high-frequency array optimized for the $CO_{linear}$ and (c) low-frequency array optimized for the $CO_{bridge}$ detection during the cathodic scan. (d-e) Evolution of IR spectra with the potential acquired on the

(d) high-frequency array optimized for the $CO_{linear}$ and (e) low-frequency array optimized for the $CO_{bridge}$ detection during the anodic scan.

## 4. Conclusion

We have developed a multi-band nanophotonic-electrochemical platform enabling enhanced simultaneous *in-situ* characterization of two adsorption configurations of CO on Pt during the electrochemical reduction of $CO_2$. Our platform provided an enhancement over conventional systems by a factor of over 40. Crucially, our platform was able to detect the $CO_{bridge}$ configuration, which could not be detected here with an unstructured Pt film. Using a straightforward and easily reproducible methodology, we numerical modeled, fabricated, and tested our platform. The $CO_{linear}$ and $CO_{bridge}$ configurations were characterized in CO saturated electrolytes, highlighting a transition from top to bridge site configurations at high cathodic potentials and demonstrating the high resolution provided by our platform. The vibrational modes of CO were confirmed *via* their typical Stark shift. Our final experimental tests focused on the characterization of the $CO_2RR$. Interestingly, we found that during the $CO_2RR$ in an alkaline environment, the $CO_{bridge}$ configuration does not play a significant role, whereas the $CO_{linear}$ was successfully detected. This finding could be attributed to the competition with the hydrogen evolution reaction in alkaline environments. We anticipate that our multi-band nanophotonic-electrochemical platform provides a new strategy to study electrochemical reactions with low coverage or transient features by providing a higher resolution than conventional systems for now limitless IR-active vibrational modes.

**Author Contributions**

M.D. and L.M.B. contributed equally to this work. The manuscript was written through contributions of all authors. All authors have given approval to the final version of the manuscript.

**Notes**

The authors declare no competing financial interest.

**Acknowledgements**

The authors thank Thomas Weber for his help with the temporal-coupled mode theory algorithms and Simon Stork for the platinum and titanium electron beam evaporation. This project was funded by the Deutsche Forschungsgemeinschaft (DFG, German Research Foundation) under grant numbers EXC 2089/1–390776260 (Germany's Excellence Strategy) and TI 1063/1 (Emmy Noether Program), the Bavarian program Solar Energies Go Hybrid (SolTech) and the Center for NanoScience (CeNS) and the European Union (ERC, METANEXT, 101078018 and NEHO, 101046329). Views and opinions expressed are however those of the author(s) only and do not necessarily reflect those of the European Union or the European Research Council Executive Agency. Neither the European Union nor the granting authority can be held responsible for them. S.A.M. additionally acknowledges the Lee-Lucas Chair in Physics and the EPSRC (EP/W017075/1).

**Abbreviations**

$CO_2RR$, $CO_2$ reduction reaction; SEIRAS, surface-enhanced IR absorption spectroscopy; ATR, attenuated total internal reflectance; OCP, open circuit potential.

**References**

(1) Wuttig, A.; Yaguchi, M.; Motobayashi, K.; Osawa, M.; Surendranath, Y. Inhibited Proton Transfer Enhances Au-Catalyzed $CO_2$-to-Fuels Selectivity. *Proc. Natl. Acad. Sci.* **2016**, *113* (32), E4585–E4593. https://doi.org/10.1073/pnas.1602984113.
(2) 3.5 In Situ Infrared Spectroelectrochemistry. In *Encyclopedia of Electrochemistry: Online*; Bard, A. J., Ed.; Wiley, 2007. https://doi.org/10.1002/9783527610426.
(3) Christensen, P. A.; Hamnett, A. Chapter 1 In-Situ Infrared Studies of the Electrode-Electrolyte Interface. In *Comprehensive Chemical Kinetics*; Elsevier, 1989; Vol. 29, pp 1–77. https://doi.org/10.1016/S0069-8040(08)70316-9.

(4) Wain, A. J.; O'Connell, M. A. Advances in Surface-Enhanced Vibrational Spectroscopy at Electrochemical Interfaces. *Adv. Phys. X* **2017**, *2* (1), 188–209. https://doi.org/10.1080/23746149.2016.1268931.

(5) Osawa, M. In-situ Surface-Enhanced Infrared Spectroscopy of the Electrode/Solution Interface. In *Advances in Electrochemical Sciences and Engineering*; Alkire, R. C., Kolb, D. M., Lipkowski, J., Ross, P. N., Eds.; Wiley, 2006; Vol. 9, pp 269–314. https://doi.org/10.1002/9783527616817.ch8.

(6) Adato, R.; Yanik, A. A.; Amsden, J. J.; Kaplan, D. L.; Omenetto, F. G.; Hong, M. K.; Erramilli, S.; Altug, H. Ultra-Sensitive Vibrational Spectroscopy of Protein Monolayers with Plasmonic Nanoantenna Arrays. *Proc. Natl. Acad. Sci.* **2009**, *106* (46), 19227–19232. https://doi.org/10.1073/pnas.0907459106.

(7) Berger, L. M.; Duportal, M.; Menezes, L. de S.; Cortés, E.; Maier, S. A.; Tittl, A.; Krischer, K. Improved In Situ Characterization of Electrochemical Interfaces Using Metasurface-Driven Surface-Enhanced IR Absorption Spectroscopy. *Adv Funct Mater* **2023**. https://doi.org/10.1002/adfm.202300411.

(8) Iwasita, T.; Nart, F. C.; Lopez, B.; Vielstich, W. On the Study of Adsorbed Species at Platinum from Methanol, Formic Acid and Reduced Carbon Dioxide via in Situ FT-ir Spectroscopy. *Electrochimica Acta* **1992**, *37* (12), 2361–2367. https://doi.org/10.1016/0013-4686(92)85133-6.

(9) Rodes, A.; Pastor, E.; Iwasita, T. Structural Effects on $CO_2$ Reduction at Pt Single-Crystal Electrodes Part 3. Pt(100) and Related Surfaces. *J. Electroanal. Chem.* **1994**, *377* (1–2), 215–225. https://doi.org/10.1016/0022-0728(94)03424-9.

(10) Arévalo, M. C.; Gomis-Bas, C.; Hahn, F.; Beden, B.; Arévalo, A.; Arvia, A. J. A Contribution to the Mechanism of "Reduced" $CO_2$ Adsorbates Electro-Oxidation from Combined Spectroelectrochemical and Voltammetric Data. *Electrochimica Acta* **1994**, *39* (6), 793–799. https://doi.org/10.1016/0013-4686(93)E0023-F.

(11) Sun, S.-G.; Zhou, Z.-Y. Surface Processes and Kinetics of $CO_2$ Reduction on Pt(100) Electrodes of Different Surface Structure in Sulfuric Acid Solutions. *Phys. Chem. Chem. Phys.* **2001**, *3* (16), 3277–3283. https://doi.org/10.1039/b100938i.

(12) Rakić, A. D.; Djurišić, A. B.; Elazar, J. M.; Majewski, M. L. Optical Properties of Metallic Films for Vertical-Cavity Optoelectronic Devices. *Appl. Opt.* **1998**, *37* (22), 5271. https://doi.org/10.1364/AO.37.005271.

(13) Haus, H. A. *Waves and Fields in Optoelectronics*; Prentice-Hall, 1984.

(14) Wang, J.; Weber, T.; Aigner, A.; Maier, S. A.; Tittl, A. Mirror-Coupled Plasmonic Bound States in the Continuum for Tunable Perfect Absorption. arXiv November 7, 2022. https://doi.org/10.48550/arXiv.2211.03673.

(15) Tittl, A.; Leitis, A.; Liu, M.; Yesilkoy, F.; Choi, D.-Y.; Neshev, D. N.; Kivshar, Y. S.; Altug, H. Imaging-Based Molecular Barcoding with Pixelated Dielectric Metasurfaces. *Science* **2018**, *360* (6393), 1105–1109. https://doi.org/10.1126/science.aas9768.

(16) Hu, H.; Weber, T.; Bienek, O.; Wester, A.; Hüttenhofer, L.; Sharp, I. D.; Maier, S. A.; Tittl, A.; Cortés, E. Catalytic Metasurfaces Empowered by Bound States in the Continuum. *ACS Nano* **2022**, *16* (8), 13057–13068. https://doi.org/10.1021/acsnano.2c05680.

(17) Kato, K.; Liu, Y.; Murakami, S.; Morita, Y.; Mori, T. Electron Beam Lithography with Negative Tone Resist for Highly Integrated Silicon Quantum Bits. *Nanotechnology* **2021**, *32* (48), 485301. https://doi.org/10.1088/1361-6528/ac201b.

(18) Hossain, M. J.; Rahman, M. M.; Jafar Sharif, Md. Preference for Low-Coordination Sites by Adsorbed CO on Small Platinum Nanoparticles. *Nanoscale Adv.* **2020**, *2* (3), 1245–1252. https://doi.org/10.1039/C9NA00499H.

(19) Silva, C. D.; Cabello, G.; Christinelli, W. A.; Pereira, E. C.; Cuesta, A. Simultaneous Time-Resolved ATR-SEIRAS and CO-Charge Displacement Experiments: The Dynamics of CO Adsorption on Polycrystalline Pt. *J. Electroanal. Chem.* **2017**, *800*, 25–31. https://doi.org/10.1016/j.jelechem.2016.10.034.

(20) McPherson, I. J.; Ash, P. A.; Jones, L.; Varambhia, A.; Jacobs, R. M. J.; Vincent, K. A. Electrochemical CO Oxidation at Platinum on Carbon Studied through Analysis of Anomalous in Situ IR Spectra. *J. Phys. Chem. C* **2017**, *121* (32), 17176–17187. https://doi.org/10.1021/acs.jpcc.7b02166.

(21) Susarrey-Arce, A.; Tiggelaar, R. M.; Gardeniers, J. G. E.; van Houselt, A.; Lefferts, L. CO Adsorption on Pt Nanoparticles in Low E-Fields Studied by ATR-IR Spectroscopy in a Microreactor. *J. Phys. Chem. C* **2015**, *119* (44), 24887–24894. https://doi.org/10.1021/acs.jpcc.5b08392.

(22) Smolinka, T.; Heinen, M.; Chen, Y. X.; Jusys, Z.; Lehnert, W.; Behm, R. J. $CO_2$ Reduction on Pt Electrocatalysts and Its Impact on $H_2$ Oxidation in $CO_2$ Containing Fuel Cell Feed Gas – A Combined in Situ Infrared Spectroscopy, Mass Spectrometry and Fuel Cell Performance Study. *Electrochimica Acta* **2005**, *50* (25–26), 5189–5199. https://doi.org/10.1016/j.electacta.2005.02.082.

(23) Farias, M. J. S.; Busó-Rogero, C.; Tanaka, A. A.; Herrero, E.; Feliu, J. M. Monitoring of CO Binding Sites on Stepped Pt Single Crystal Electrodes in Alkaline Solutions by in Situ FTIR Spectroscopy. *Langmuir* **2020**, *36* (3), 704–714. https://doi.org/10.1021/acs.langmuir.9b02928.

(24) Blizanac, B. B.; Lucas, C. A.; Gallagher, M. E.; Arenz, M.; Ross, P. N.; Marković, N. M. Anion Adsorption, CO Oxidation, and Oxygen Reduction Reaction on a Au(100) Surface: The PH Effect. *J. Phys. Chem. B* **2004**, *108* (2), 625–634. https://doi.org/10.1021/jp036483l.

(25) Mayet, N.; Servat, K.; Kokoh, K. B.; Napporn, T. W. Electrochemical Oxidation of Carbon Monoxide on Unsupported Gold Nanospheres in Alkaline Medium. *Electrocatalysis* **2021**, *12* (1), 26–35. https://doi.org/10.1007/s12678-020-00626-7.

(26) Villegas, I.; Weaver, M. J. Carbon Monoxide Adlayer Structures on Platinum (111) Electrodes: A Synergy between *In-situ* Scanning Tunneling Microscopy and Infrared Spectroscopy. *J. Chem. Phys.* **1994**, *101* (2), 1648–1660. https://doi.org/10.1063/1.467786.

(27) Nakamura, M.; Ogasawara, H.; Inukai, J.; Ito, M. CO Migration on Pt(100) and Pt(11 1 1) Surfaces Studied by Time Resolved Infrared Reflection-Absorption Spectroscopy. *Surf. Sci.* **1993**, *283* (1–3), 248–254. https://doi.org/10.1016/0039-6028(93)90989-W.

(28) Katayama, Y.; Giordano, L.; Rao, R. R.; Hwang, J.; Muroyama, H.; Matsui, T.; Eguchi, K.; Shao-Horn, Y. Surface (Electro)Chemistry of $CO_2$ on Pt Surface: An *in Situ* Surface-Enhanced Infrared Absorption Spectroscopy Study. *J. Phys. Chem. C* **2018**, *122* (23), 12341–12349. https://doi.org/10.1021/acs.jpcc.8b03556.

(29) Dunwell, M. Surface Enhanced Spectroscopic Investigations of Adsorption of Cations on Electrochemical Interfaces. *Phys Chem Chem Phys* **2017**, *19* (971). https://doi.org/10.1039/c6cp07207k.

(30) Samjeské, G.; Komatsu, K.; Osawa, M. Dynamics of CO Oxidation on a Polycrystalline Platinum Electrode: A Time-Resolved Infrared Study. *J. Phys. Chem. C* **2009**, *113* (23), 10222–10228. https://doi.org/10.1021/jp900582c.

(31) Stamenkovic, V.; Chou, K. C.; Somorjai, G. A.; Ross, P. N.; Markovic, N. M. Vibrational Properties of CO at the Pt(111)−Solution Interface: The Anomalous Stark-Tuning Slope. *J. Phys. Chem. B* **2005**, *109* (2), 678–680. https://doi.org/10.1021/jp044802i.

(32) Chang, S.-C.; Weaver, M. J. In-Situ Infrared Spectroscopy of CO Adsorbed at Ordered Pt(110)-Aqueous Interfaces. *Surf. Sci.* **1990**, *230*, 222–236. https://doi.org/10.1016/0039-6028(90)90030-C.

(33) Deshlahra, P.; Conway, J.; Wolf, E. E.; Schneider, W. F. Influence of Dipole–Dipole Interactions on Coverage-Dependent Adsorption: CO and NO on Pt(111). *Langmuir* **2012**, *28* (22), 8408–8417. https://doi.org/10.1021/la300975s.

(34) Lambert, D. K. Vibrational Stark Effect of Adsorbates at Electrochemical Interfaces. *Electrochimica Acta* **1996**, *41* (5), 623–630. https://doi.org/10.1016/0013-4686(95)00349-5.

(35) Ringe, S.; Clark, E. L.; Resasco, J.; Walton, A.; Seger, B.; Bell, A. T.; Chan, K. Understanding Cation Effects in Electrochemical $CO_2$ Reduction. *Energy Environ. Sci.* **2019**, *12* (10), 3001–3014. https://doi.org/10.1039/C9EE01341E.

(36) Wang, H.; Tobin, R. G.; Lambert, D. K. Coadsorption of Hydrogen and CO on Pt(335): Structure and Vibrational Stark Effect. *J. Chem. Phys.* **1994**, *101* (5), 4277–4287. https://doi.org/10.1063/1.467478.

(37) Fan, S. Sharp Asymmetric Line Shapes in Side-Coupled Waveguide-Cavity Systems. *Appl. Phys. Lett.* **2002**, *80* (6), 908–910. https://doi.org/10.1063/1.1448174.

(38) Alavi, A.; Hu, P.; Deutsch, T.; Silvestrelli, P. L.; Hutter, J. CO Oxidation on Pt(111): An *Ab Initio* Density Functional Theory Study. *Phys. Rev. Lett.* **1998**, *80* (16), 3650–3653. https://doi.org/10.1103/PhysRevLett.80.3650.

(39) Rheinländer, P.; Henning, S.; Herranz, J.; Gasteiger, H. A. Comparing Hydrogen Oxidation and Evolution Reaction Kinetics on Polycrystalline Platinum in 0.1 M and 1 M KOH. *ECS Trans.* **2013**, *50* (2), 2163–2174. https://doi.org/10.1149/05002.2163ecst.

(40) Giner, J. Electrochemical Reduction of $CO_2$ on Platinum Electrodes in Acid Solutions. *Electrochimica Acta* **1963**, *8* (11), 857–865. https://doi.org/10.1016/0013-4686(63)80054-7.

(41) Kamat, G. A.; Zamora Zeledón, J. A.; Gunasooriya, G. T. K. K.; Dull, S. M.; Perryman, J. T.; Nørskov, J. K.; Stevens, M. B.; Jaramillo, T. F. Acid Anion Electrolyte Effects on Platinum for Oxygen and Hydrogen Electrocatalysis. *Commun. Chem.* **2022**, *5* (1), 20. https://doi.org/10.1038/s42004-022-00635-1.

(42) Spendelow, J. S.; Goodpaster, J. D.; Kenis, P. J. A.; Wieckowski, A. Mechanism of CO Oxidation on Pt(111) in Alkaline Media. *J. Phys. Chem. B* **2006**, *110* (19), 9545–9555. https://doi.org/10.1021/jp060100c.

(43) Watanabe, S.; Inukai, J.; Ito, M. Coverage and Potential Dependent CO Adsorption on Pt(111), (711) and (100) Electrode Surfaces Studied by Infrared Reflection Absorption Spectroscopy. *Surf. Sci.* **1993**, *293*, 1–9. https://doi.org/10.1016/0039-6028(93)90237-E.

(44) Sheng, W.; Gasteiger, H. A.; Shao-Horn, Y. Hydrogen Oxidation and Evolution Reaction Kinetics on Platinum: Acid vs Alkaline Electrolytes. *J. Electrochem. Soc.* **2010**, *157* (11), B1529. https://doi.org/10.1149/1.3483106.

(45) Taji, Y.; Zagalskaya, A.; Evazzade, I.; Watzele, S.; Song, K.-T.; Xue, S.; Schott, C.; Garlyyev, B.; Alexandrov, V.; Gubanova, E.; Bandarenka, A. S. Alkali Metal Cations Change the Hydrogen Evolution Reaction Mechanisms at Pt Electrodes in Alkaline Media. *Nano Mater. Sci.* **2022**, S2589965122000514. https://doi.org/10.1016/j.nanoms.2022.09.003.

(46) Nørskov, J. K.; Bligaard, T.; Logadottir, A.; Kitchin, J. R.; Chen, J. G.; Pandelov, S.; Stimming, U. Trends in the Exchange Current for Hydrogen Evolution. *J. Electrochem. Soc.* **2005**, *152* (3), J23–J26. http://dx.doi.org/10.1149/1.1856988.